\documentstyle[12pt]{article}
\newcommand{\nc}{\newcommand}
\nc{\renc}{\renewcommand}

\nc{\half}{{\textstyle{1\over2}}}

\nc{\etal}{\mbox{\it et al. }}
\nc{\ie}{{\it i.e.}}
\nc{\eg}{{\it e.g.}}

\renc{\thefootnote}{\arabic{footnote}}
\nc{\capt}[1]{{\bf Figure.} {\small\sl #1}}

\nc{\eqs}[2]{\mbox{Eqs.~(\ref{#1},\,\ref{#2})}}
\nc{\eq}[1]{\mbox{Eq.~(\ref{#1})}}

\nc{\figs}[2]{\mbox{Figs.~(\ref{#1},\,\ref{#2})}}
\nc{\fig}[1]{\mbox{Fig~.(\ref{#1})}}

\nc{\tag}[1]{\label{#1} \marginpar{{\footnotesize #1}}}
\nc{\mtag}[1]{\label{#1} \mbox{\marginpar{{\footnotesize #1}}}}
\renc{\baselinestretch}{1.2}
\newlength{\overeqskip}
\newlength{\undereqskip}
\nc{\be}[1]{\begin{equation} \mbox{$\label{#1}$}}
\nc{\bea}[1]{\begin{eqnarray} \mbox{$\label{#1}$}}
\nc{\Section}[2]{\section{#2}\label{#1}}
\nc{\Bibitem}[1]{\bibitem{#1}}
\nc{\Label}[1]{\label{#1}}

\nc{\eea}{\vspace{\undereqskip}\end{eqnarray}}
\nc{\ee}{\vspace{\undereqskip}\end{equation}}
\nc{\bdm}{\begin{displaymath}}
\nc{\edm}{\end{displaymath}}
\nc{\dpsty}{\displaystyle}
\nc{\bc}{\begin{center}}
\nc{\ec}{\end{center}}
\nc{\ba}{\begin{array}}
\nc{\ea}{\end{array}}
\nc{\bab}{\begin{abstract}}
\nc{\eab}{\end{abstract}}
\nc{\btab}{\begin{tabular}}
\nc{\etab}{\end{tabular}}
\nc{\bit}{\begin{itemize}}
\nc{\eit}{\end{itemize}}
\nc{\ben}{\begin{enumerate}}
\nc{\een}{\end{enumerate}}
\nc{\bfig}{\begin{figure}}
\nc{\efig}{\end{figure}}
\nc{\arreq}{&\!=\!&}
\nc{\arrmi}{&\!-\!&}
\nc{\arrpl}{&\!+\!&}
\nc{\arrap}{&\!\!\!\approx\!\!\!&}
\nc{\non}{\nonumber\\*}
\nc{\align}{\!\!\!\!\!\!\!\!&&}

\def\lsim{\; \raise0.3ex\hbox{$<$\kern-0.75em
      \raise-1.1ex\hbox{$\sim$}}\; }
\def\gsim{\; \raise0.3ex\hbox{$>$\kern-0.75em
      \raise-1.1ex\hbox{$\sim$}}\; }
\nc{\DOT}{\hspace{-0.08in}{\bf .}\hspace{0.1in}}
\nc{\Laada}{\hbox {$\sqcap$ \kern -1em $\sqcup$}}
\nc\loota{{\scriptstyle\sqcap\kern-0.55em\hbox{$\scriptstyle\sqcup$}}}
\nc\Loota{{\sqcap\kern-0.65em\hbox{$\sqcup$}}}
\nc\laada{\Loota}
\nc{\qed}{\hskip 3em \hbox{\BOX} \vskip 2ex}

\nc{\real}{{\rm I \! R}}
\nc{\Z}{{\sf Z \!\!\! Z}}
\nc{\complex}{{\rm C\!\!\! {\sf I}\,\,}}
\def\bigid{\leavevmode\hbox{\small1\kern-3.8pt\normalsize1}}
\def\id{\leavevmode\hbox{\small1\kern-3.3pt\normalsize1}}
\nc{\slask}{\!\!\!/}
\nc{\bis}{{\prime\prime}}
\nc{\pa}{\partial}
\nc{\na}{\nabla}
\nc{\ra}{\rangle}
\nc{\la}{\langle}
\nc{\goto}{\rightarrow}
\nc{\swap}{\leftrightarrow}

\nc{\EE}[1]{ \mbox{$\cdot10^{#1}$} }
\nc{\abs}[1]{\left|#1\right|}
\nc{\at}[2]{\left.#1\right|_{#2}}
\nc{\norm}[1]{\|#1\|}
\nc{\abscut}[2]{\Abs{#1}_{\scriptscriptstyle#2}}
\nc{\vek}[1]{{\rm\bf #1}}
\nc{\integral}[2]{\int\limits_{#1}^{#2}}
\nc{\inv}[1]{\frac{1}{#1}}
\nc{\dd}[2]{{{\partial #1}\over{\partial #2}}}
\nc{\ddd}[2]{{{{\partial}^2 #1}\over{\partial {#2}^2}}}
\nc{\dddd}[3]{{{{\partial}^2 #1}\over
        {\partial #2 \partial #3}}}
\nc{\dder}[2]{{{d #1}\over{d #2}}}
\nc{\ddder}[2]{{{d^2 #1}\over{d {#2}^2}}}
\nc{\dddder}[3]{{d^2 #1}\over
        {d #2 d #3}}
\nc{\dx}[1]{d\,^{#1}x}
\nc{\dy}[1]{d\,^{#1}y}
\nc{\dz}[1]{d\,^{#1}z}
\nc{\dl}[1]{\frac{d\,^{#1}l}{(2\pi)^{#1}}}
\nc{\dk}[1]{\frac{d\,^{#1}k}{(2\pi)^{#1}}}
\nc{\dq}[1]{\frac{d\,^{#1}q}{(2\pi)^{#1}}}

\nc{\cc}{\mbox{$c.c.$ }}
\nc{\hc}{\mbox{$h.c.$ }}
\nc{\cf}{cf.\ }
\nc{\erfc}{{\rm erfc}}
\nc{\Tr}{{\rm Tr\,}}
\nc{\tr}{{\rm tr\,}}
\nc{\pol}{{\rm pol}}
\nc{\sign}{{\rm sign}}
\nc{\bfT}{{\bf T }}

\nc{\cA}{{\cal A}}
\nc{\cB}{{\cal B}}
\nc{\cD}{{\cal D}}
\nc{\cE}{{\cal E}}
\nc{\cG}{{\cal G}}
\nc{\cH}{{\cal H}}
\nc{\cL}{{\cal L}}
\nc{\cO}{{\cal O}}
\nc{\cT}{{\cal T}}
\nc{\cN}{{\cal N}}
\nc{\rvac}[1]{|{\cal O}#1\rangle}
\nc{\lvac}[1]{\langle{\cal O}#1|}
\nc{\rvacb}[1]{|{\cal O}_\beta #1\rangle}
\nc{\lvacb}[1]{\langle{\cal O}_\beta #1 |}
\nc{\bb}{\bar{\beta}}
\nc{\bt}{\tilde{\beta}}
\nc{\ctH}{\tilde{\cal H}}
\nc{\chH}{\hat{\cal H}}
\nc{\1}{\aa}
\nc{\2}{\"{a}}
\nc{\3}{\"{o}}
\nc{\4}{\AA}
\nc{\5}{\"{A}}
\nc{\6}{\"{O}}
\nc{\al}{\alpha}
\nc{\g}{\gamma}
\nc{\Del}{\Delta}
\nc{\e}{\epsilon}
\nc{\eps}{\epsilon}
\nc{\lam}{\lambda}
\nc{\om}{\omega}
\nc{\Om}{\Omega}
\nc{\ve}{\varepsilon}
\nc{\mn}{{\mu\nu}}
\nc{\vp}{\varphi}

\nc{\advp}[3]{{\it  Adv.\ in\ Phys.\ }{{\bf #1} {(#2)} {#3}}}
\nc{\annp}[3]{{\it  Ann.\ Phys.\ (N.Y.)\ }{{\bf #1} {(#2)} {#3}}}
\nc{\apl}[3]{{\it  Appl. Phys. Lett. }{{\bf #1} {(#2)} {#3}}}
\nc{\apj}[3]{{\it  Ap.\ J.\ }{{\bf #1} {(#2)} {#3}}}
\nc{\apjl}[3]{{\it  Ap.\ J.\ Lett.\ }{{\bf #1} {(#2)} {#3}}}
\nc{\app}[3]{{\it Astropart.\ Phys.\ }{{\bf #1} {(#2)} {#3}}}
\nc{\cmp}[3]{{\it  Comm.\ Math.\ Phys.\ }{{ \bf #1} {(#2)} {#3}}}
\nc{\cqg}[3]{{\it  Class.\ Quant.\ Grav.\ }{{\bf #1} {(#2)} {#3}}}
\nc{\epl}[3]{{\it  Europhys.\ Lett.\ }{{\bf #1} {(#2)} {#3}}}
\nc{\ijmp}[3]{{\it Int.\ J.\ Mod.\ Phys.\ }{{\bf #1} {(#2)} {#3}}}
\nc{\ijtp}[3]{{\it Int.\ J.\ Theor.\ Phys.\ }{{\bf #1} {(#2)} {#3}}}
\nc{\jmp}[3]{{\it  J.\ Math.\ Phys.\ }{{ \bf #1} {(#2)} {#3}}}
\nc{\jpa}[3]{{\it  J.\ Phys.\ A\ }{{\bf #1} {(#2)} {#3}}}
\nc{\jpc}[3]{{\it  J.\ Phys.\ C\ }{{\bf #1} {(#2)} {#3}}}
\nc{\jap}[3]{{\it J.\ Appl.\ Phys.\ }{{\bf #1} {(#2)} {#3}}}
\nc{\jpsj}[3]{{\it J.\ Phys.\ Soc.\ Japan\ }{{\bf #1} {(#2)} {#3}}}
\nc{\lmp}[3]{{\it Lett.\ Math.\ Phys.\ }{{\bf #1} {(#2)} {#3}}}
\nc{\mpl}[3]{{\it  Mod.\ Phys.\ Lett.\ }{{\bf #1} {(#2)} {#3}}}
\nc{\ncim}[3]{{\it  Nuov.\ Cim.\ }{{\bf #1} {(#2)} {#3}}}
\nc{\np}[3]{{\it  Nucl.\ Phys.\ }{{\bf #1} {(#2)} {#3}}}
\nc{\pr}[3]{{\it Phys.\ Rev.\ }{{\bf #1} {(#2)} {#3}}}
\nc{\pra}[3]{{\it  Phys.\ Rev.\ A\ }{{\bf #1} {(#2)} {#3}}}
\nc{\prb}[3]{{\it  Phys.\ Rev.\ B\ }{{{\bf #1} {(#2)} {#3}}}}
\nc{\prc}[3]{{\it  Phys.\ Rev.\ C\ }{{\bf #1} {(#2)} {#3}}}
\nc{\prd}[3]{{\it  Phys.\ Rev.\ D\ }{{\bf #1} {(#2)} {#3}}}
\nc{\prl}[3]{{\it Phys.\ Rev.\ Lett.\ }{{\bf #1} {(#2)} {#3}}}
\nc{\pl}[3]{{\it  Phys.\ Lett.\ }{{\bf #1} {(#2)} {#3}}}
\nc{\prep}[3]{{\it Phys.\ Rep.\ }{{\bf #1} {(#2)} {#3}}}
\nc{\prsl}[3]{{\it Proc.\ R.\ Soc.\ London\ }{{\bf #1} {(#2)} {#3}}}
\nc{\ptp}[3]{{\it  Prog.\ Theor.\ Phys.\ }{{\bf #1} {(#2)} {#3}}}
\nc{\ptps}[3]{{\it  Prog\ Theor.\ Phys.\ suppl.\ }{{\bf #1} {(#2)} {#3}}}
\nc{\physa}[3]{{\it  Physica\ A\ }{{\bf #1} {(#2)} {#3}}}
\nc{\physb}[3]{{\it  Physica\ B\ }{{\bf #1} {(#2)} {#3}}}
\nc{\phys}[3]{{\it Physica\ }{{\bf #1} {(#2)} {#3}}}
\nc{\rmp}[3]{{\it  Rev.\ Mod.\ Phys.\ }{{\bf #1} {(#2)} {#3}}}
\nc{\rpp}[3]{{\it Rep.\ Prog.\ Phys.\ }{{\bf #1} {(#2)} {#3}}}
\nc{\sjnp}[3]{{\it Sov.\ J.\ Nucl.\ Phys.\ }{{\bf #1} {(#2)} {#3}}}
\nc{\spjetp}[3]{{\it Sov.\ Phys.\ JETP\ }{{\bf #1} {(#2)} {#3}}}
\nc{\yf}[3]{{\it Yad.\ Fiz.\ }{{\bf #1} {(#2)} {#3}}}
\nc{\zetp}[3]{{\it Zh.\ Eksp.\ Teor.\ Fiz.\  }{{\bf #1}  {(#2)} {#3}}}
\nc{\zp}[3]{{\it Z.\ Phys.\ }{{\bf #1} {(#2)} {#3}}}
\nc{\ibid}[3]{{\sl ibid.\ }{{\bf #1} {#2} {#3}}}
\nc{\rf}[1]{(\ref{#1})}
\nc{\nn}{\nonumber \\*}
\nc{\bfB}{\bf{B}}
\nc{\bfv}{\bf{v}}
\nc{\bfx}{\bf{x}}
\nc{\bfy}{\bf{y}}
\nc{\vx}{\vec{x}}
\nc{\vy}{\vec{y}}
\nc{\oB}{\overline{B}}
\nc{\oI}{\overline{I}}
\nc{\oR}{\overline{R}}
\nc{\rar}{\rightarrow}
\nc{\ti}{\times}
\nc{\slsh}{\hskip-5pt/}
\nc{\sm}{Standard~Model~}
\nc{\MP}{M_{\rm Pl}}
\nc{\tp}{t_{\rm Pl}}
\nc{\ave}{\bar{E}}

\nc{\eff}{{\rm eff}}
\nc{\kk}{\vek{k}}
\nc{\pp}{{\rm p}}
\nc{\ga}{g_{a\gamma}}
\nc{\vv}{\\}
\nc{\eee}{{\bf E}}
\nc{\bbb}{{\bf B}}
\nc{\qcd}{T_{\rm QCD}}
\nc{\G}{\rm \ G}
\def\vec#1{{\bf #1}}
\begin{document}
{\title{\vskip-2truecm{\hfill {{\small HIP-1997-56/TH\\
        }}\vskip 1truecm}
{\bf Testing exotic neutrino-neutrino interactions with AGN neutrinos}}


{\author{
{\sc Petteri Ker\"anen$^{1}$ 
}\\
{\sl\small Department of Physics, P.O. Box 9,
FIN-00014 University of Helsinki,
Finland}
}
\maketitle
\vspace{2cm}
\begin{abstract}
\noindent
We propose a test for non-standard neutrino-neutrino interactions by using ultrahigh energy AGN neutrinos. Such interactions would influence the AGN neutrino flux due to collisions with cosmic background neutrinos. For typical AGN neutrinos we obtain an upper limit for the coupling constant $g<6.4\cdot 10^{-3}$ if the mediator is light and $g/(M_X/{\rm MeV}) <0.013$ if the mediator is heavy. We compare our results with constraints from other phenomena previously considered.
\end{abstract}
\vfil
\footnoterule
{\small $^{1}$petteri.keranen@helsinki.fi\vskip-1pt\noindent}
\thispagestyle{empty}
\newpage
\setcounter{page}{1}


\noindent {\it 1. Introduction.} Active galactic nuclei (AGN) are found to be the origin of the highest energy photons detected in gamma ray observatories. The production of these photons is believed to be accompanied by a production of other high-energy cosmic rays, like protons, electrons and neutrinos~\cite{Gaisser}. The weakly interacting neutrinos can carry us first-hand information of the processes taking place in the source and of the source itself, and they can reach us from more distant AGN. On the other hand, the AGN neutrinos can be useful in studying the properties of neutrinos themselves. The high-energies not achievable in laboratory experiments or in other astrophysical environments, combined with long distances between the source and the detection point, may reveal new important information of mixing, masses and possible new interaction forms of neutrinos. 

In this paper we will study the possibility of testing non-standard neutrino-neutrino interactions with high-energy neutrinos from AGN. If such interactions exist, the AGN neutrinos will undergo extra collision processes with neutrinos in cosmic background in their way from the source to detectors. This will affect their measured flux.  We will derive an upper limit for the effective strength of these interactions that could be reached by measuring the flux in the new neutrino telescopes such as AMANDA~\cite{amanda}, Nestor~\cite{nestor}, Baikal~\cite{baikal} and ANTARES~\cite{antares}. 
The energies of the extragalactic neutrinos we are interested in are typically from some hundreds of GeV up to a few EeV:s. The typical 
length scale is several hundreds of megaparsecs which is the distance to the neighbouring 
active galaxies. The new neutrino telescopes can detect very high energy neutrinos with quite a good angular resolution, in the future even allowing one to connect observed neutrinos with given AGN sources despite the atmospheric neutrino background~\cite{antares2}. 

In the standard model neutrinos
interact with other neutrinos only via $Z$-boson exchange. Neutrino-neutrino interactions are, however, hard to test, and it is not excluded that there are substantial non-standard "secret" forces between neutrinos. We will consider the possibility of testing such secret forces using AGN neutrinos. To be specific, we will assume that these interactions are of the form ($i=e,\mu,\tau $) 
\bea{gcoupl} 
g \bar{\nu}_i \gamma_{\mu} \nu_i X^{\mu} ,
\eea
where $X$ is a spin-one boson and $g$ is the coupling constant 
of the interaction. We will here assume that the coupling is similar between different neutrinos and that this new vectorial interaction 
does not couple $X$ boson with any other particles than neutrinos.
The effective Hamiltonian of this interaction in the low energy approximation is
\bea{FVHam}
{\cal H} = F_V (\bar\nu \gamma_\alpha \nu ) (\bar\nu \gamma^\alpha \nu ),
\eea
where $F_V=g^2/M_X^2$ is the effective coupling constant of the interaction. In the order of magnitude level our results are obviously valid for a wider class of models than just for the one defined in~\eq{gcoupl} and~\eq{FVHam}.

Laboratory constraints for the non-standard interactions of the form given above have been discussed in~\cite{Bardin} -~\cite{Bilenky}. The most stringent one comes from measuring the
invisible width of the $Z$-boson at LEP~\cite{Bilenky}:
\bea{FVlimit}
F_V < 4 \times 10^2 G_F.
\eea
Here the number of 
neutrino species have been used to set a limit on the non-standard four neutrino coupling resulting in the cascade decay $Z\rightarrow \nu \bar\nu \rightarrow \nu \bar\nu \nu \bar\nu$.

Constraints from cosmology and astrophysics have been considered in~\cite{Manohar} -~\cite{Masso}. One of these limits~\cite{Masso} is even stricter than the laboratory limit (\ref{FVlimit}), but it is more model dependent, as will be discussed below. These constraints come from SN1987A~\cite{Manohar,Kolb} and nucleosynthesis considerations~\cite{Masso}. 

It is interesting to study whether one could improve the present bounds by measuring the flux of AGN neutrinos. We will show that indeed for certain values of the $X$ boson mass more stringent limits will be achievable.  
\\

\noindent {\it 2. UHE neutrino sources.} In the litterature two types of mechanisms for neutrino production in AGN have been discussed, spherical accretion mechanism and blazars producing jets (for a review, see~\cite{Gaisser}). 

In the spherical accretion mechanism it is believed that in the center of an active galaxy a supermassive black hole creates an  accretion disk of materia falling into the black hole. In this disk a shock is formed that accelerates protons~\cite{ProKaz},~\cite{KazEll} to very high energies. Neutrinos are produced in pion photoproduction processes and in $pp$-collisions, the most important reaction chain being $p\gamma \rightarrow n\pi^+\rightarrow\mu^+\nu_{\mu} \rightarrow e^+\nu_e\bar{\nu}_{\mu}$~\cite{Stecker}. Different models have been presented~\cite{SzaPro,SteDoSaSo,BeRuSi,SiBe} for estimating the fluxes of nucleons and neutrinos starting from the observed X ray flux.

It is argued that in blazars protons are accelerated in jets~\cite{MaBie} - \cite{Mann} by the first order shock acceleration mechanism, and neutrinos are then created in the decays of photoproduced pions. The flux of neutrinos can be calculated from the observed photon spectrum~\cite{BieStr}. 

In both cases of AGN tau neutrinos are produced only negligible amounts, but they might appear due to mixing. In case the secret interactions exist, the possible flux of tau neutrinos will be reduced by the same amount as that of $\nu_e$ and $\nu_{\mu}$.

The detected flux of neutrinos from a single source is estimated to be from a few to even some hundreds of events per year in a square kilometer neutrino detector~\cite{antares2}. However, the statistics will improve with time and when the plans to build several new kilometer scale detectors are to come reality. Also, the higher is the neutrino energy, the better is its detection rate in a neutrino telescope compared with that of the background atmospheric neutrinos. Moreover, one can increase the statistics by integrating the flux of all AGN and blazars of the whole known Universe. The integrated neutrino flux is expected to be well above the background flux of atmospheric neutrinos~\cite{Gaisser}. 

To obtain a conservative limit on the strength of the secret interactions we approximate the distance of AGN and blazar sources to be 500 Mpc, which is less than the actual distance of the majority of them. More stringent limits will be obtained by statistically indentifying the individual sources of the observed neutrinos and using the corresponding actual distance. 
\\


\noindent {\it 3. An order of magnitude estimate.} Constraints on the secret neutrino-neutrino interactions will be obtained from the possible depletion of the AGN neutrino flux. To obtain an order of magnitude estimate we will compare the mean free path $\lambda$ of AGN neutrinos associated with these interactions and the distance $D$ to the source with each other. An upper limit for the strength of the force is set by the condition $\lambda^{-1}D<1$. 

Assuming that the background neutrinos are relativistic Dirac particles and that the interaction is of the vector type (\ref{gcoupl}), 
one can estimate the cross section of an ultrarelativistic neutrino colliding with a cosmic background neutrino as 
\bea{xsec}
\sigma = \frac{g^4 s}{(s - M^2_X)^2} ,  
\eea
where $s\propto 2E_1T$, $E_1$ is the energy of 
the AGN neutrino and $T=1.9 \ {\rm K} = 1.64 \cdot 10^{-4} \ {\rm eV}$ is the temperature 
of the background neutrino. Here we have not taken into account the thermal distribution of the neutrino background. If $\sqrt{s} \gg M_X$, then the mass of the mediator can be neglected
and the cross section of~\eq{xsec} can be approximated as $\sigma = g^4 / s$. On the other hand, if 
$\sqrt{s} \ll M_X$, then the cross section can be written as $\sigma = g^4 s/ M_X^4$. 

The mean free path of the high energy neutrino is 
$\lambda = 1/ (n_{\nu} \sigma )$,
where $n_{\nu}$ is 
the number density of the background neutrinos,
which for one neutrino family is $n_{\nu_i} =  n_{\bar{\nu}_i} = 55 \ {\rm cm}^{-3}$. For representative values of incoming neutrino energy of $E_1 = 10^{15} \ {\rm eV}$ and distance to the source of $D = 500 \ {\rm Mpc} = 1.5 \cdot 10^{27} \ {\rm cm}$, the condition $\lambda^{-1} D <1$ leads, assuming the neutrino flux to be according to the theoretical models, for the coupling constant an upper bound of 
\bea{masslessg}
g< (s/( n_{\nu} \lambda) )^{1/4} \simeq 0.01
\eea 
in the case $\sqrt{s} \gg M_X$ and
\bea{massiveg}
g/(M_X/{\rm MeV}) < (n_{\nu} \lambda s)^{-1/4} \simeq 0.02
\eea
in the case $\sqrt{s} \ll M_X$. 

In the next section we will define this crude estimate by adding up the cross sections of all relevant processes and also taking into account the thermal distribution of the background neutrinos. A similar analyses for supernova SN1987A neutrinos is presented by Kolb and Turner in~\cite{Kolb}.
\\


\noindent{\it 4. Thermal analysis.}
We can write the Boltzmann equation of high energy neutrinos $\nu_1$ with phase space distribution $f'$ and background neutrinos $\nu_2$, assumed to be relativistic, with Fermi-Dirac distribution $f$ in the form
\bea{Boeq1}
- \frac{1}{f'} \frac{ d f'(\vec{p}_1) }{ d t } = \int \frac{d^3 \vec{p}_2}{(2\pi)^3}
                                                f(\vec{p}_2) |\Delta \vec{v}|
                                                \sigma (s) ,
\eea
where $s=(p_1 + p_2)^2$ and $\Delta \vec{v} = \vec{v}_2 - \vec{v}_1$. Here we have omitted the usual Pauli blocking factors as there is no Fermi degeneracy in the cosmic background. The cross sections $\sigma (s)$ for each different process 
are given in the table~\ref{tabliska} below, 
where we have taken as an example high energy electron neutrino scatterings with background neutrinos or antineutrinos and annihilations producing neutrino-antineutrino pairs. The numerical values of the cross sections on the last three rows are multiplied with two since there are two possible final states. Muon and tau neutrinos and antineutrinos undergo the corresponding reactions. 

In terms of $y=t|\vec{v}_1|$, where $t$ is the flight time from the source to the detector and
$y$ the distance, the Boltzmann equation reads
\bea{Boeq2}
- \frac{1}{f'} \frac{ d f' }{ d y } 
                                  \simeq \frac{ \sqrt{2} }{ 4\pi^2 } \int_0^{\infty} dE_2 E_2^2 f(E_2)                                                                                  \int_{-1}^1 dz (1-z)^{1/2} \sigma (s) .
%
\eea
The right-hand side of~\eq{Boeq2} is the inverse of the mean free path, $\lambda^{-1}$. We have here approximated that $s \simeq 2 E_1 E_2 ( 1 - z)$ and $|\Delta \vec{v}| \simeq \sqrt{2(1 + z )}$,
where $z=\cos\theta$ and $\theta$ is the angle between the initial state momenta. 

The total cross section of the reactions an AGN neutrino undergoes is of the form $\sigma=a/s$ and
$\sigma=as/M_X^4$ in the case $\sqrt{s} \gg M_X$ and $\sqrt{s} \ll M_X$, where the
dimensionless constant $a$ will be calculated numerically below in table~\ref{tabliska}. The mean free path of the AGN neutrinos is given by the expression
\bea{lmless1}
\lambda^{-1} = \frac{a}{E_1}  \frac{ \int_0^{\infty} dE_2 E_2 f(E_2) } 
                                   {\int_0^{\infty} dE_2 E_2^2 f(E_2) } n_2
             = \frac{a}{E_1 T} \frac{\pi^2}{12 \thinspace \zeta (3)} n_2
\eea
in the case $\sqrt{s} \gg M_X$ and
\bea{lmsive1}
\lambda^{-1} =  \frac{16}{5}
                \frac{aE_1}{M_X^4}  \frac{ \int_0^{\infty} dE_2 E_2^3 f(E_2) }
                                       {\int_0^{\infty} dE_2 E_2^2 f(E_2) } n_2
             =  \frac{16aE_1 T }{5M_X^4} \frac{7\pi^4}{180 \thinspace \zeta (3)} n_2 .
\eea
in the case $\sqrt{s} \ll M_X$, where $n_2$ is the number density of the cosmic background neutrinos. 


\begin{table}

\begin{tabular}{lclcl}
\hline \hline 
process & $(d\sigma /dt) (8\pi s^2 /g^4)$ & $\sigma s/g^4 $
& $(d\sigma /dt) (8\pi M^4 /g^4)$ & $\sigma  M^4/g^4 s $\\ \hline 

$\nu_e \nu_e           \rightarrow \nu_e \nu_e                          $ & ${ \scriptstyle 2(1+\frac{s^2}{t^2} + \frac{s^2}{(s+t)^2} ) }   $& 1.478 & ${ \scriptstyle 4s^2 + t^2 + (s+t)^2 } $& 0.186 \\

$\nu_e \bar{\nu}_e     \rightarrow \nu_e \bar{\nu}_e                    $ & ${ \scriptstyle 2(1+\frac{(s+t)^2}{t^2} +\frac{(s+t)^2}{s^2} )}$&   0.512 & ${ \scriptstyle 4(s+t)^2 + s^2 + t^2 } $&  0.106 \\

$\nu_e \nu_{\mu ,\tau} \rightarrow \nu_e \nu_{\mu ,\tau}                $ & ${ \scriptstyle \frac{s^2}{t^2} + \frac{(s+t)^2}{t^2}  }       $& $2\times  0.569$ & ${ \scriptstyle (s+t)^2 + s^2  }      $& $2\times 0.053$ \\

$\nu_e \bar{\nu}_{\mu ,\tau} \rightarrow \nu_e \bar{\nu}_{\mu ,\tau}                $ & ${ \scriptstyle \frac{s^2}{t^2} + \frac{(s+t)^2}{t^2}  }       $& $2\times  0.569$ & ${ \scriptstyle (s+t)^2 + s^2  }      $& $2\times 0.053$ \\

$\nu_e \bar{\nu}_e     \rightarrow \nu_{\mu ,\tau} \bar{\nu}_{\mu ,\tau}$ & ${ \scriptstyle \frac{t^2}{s^2} + \frac{(s+t)^2}{s^2}  }       $& $2\times  0.027$ & ${ \scriptstyle (s+t)^2 + t^2  } $ & $ 2\times 0.027$\\
\hline \hline
constant $a$ & & $  4.32g^4$ & & $0.558g^4$ \\
\hline \hline
\end{tabular}
\caption[rosessit]{Processes and cross sections. The differential and integrated numerical values of the cross sections of the processes are listed in the second and third column in the case $\sqrt{s} \gg M_X$ and in the last two columns in the case $\sqrt{s} \ll M_X$. } 
\label{tabliska}

\end{table}

As can be seen from the table~\ref{tabliska}, some of the cross sections are divergent when $t=-s$ or $t=0$. In the case the cross section has a divergency at $t=-s$, the integration is started from $t=-(1-\epsilon )s$ where $\epsilon$ is set to be 0.1. If there is a divergency at $t=0$, the integration is performed up to $t=-\epsilon s$.  This procedure will not substantially affect the flux of high energy neutrinos: when $t\rightarrow 0$, the collision is elastic and no reaction occurs, and when $t\rightarrow -s$, the final state neutrino has the same momentum as the initial state neutrino. 

Requiring now $\lambda^{-1} D<1$, where $D$ is the distance to the source we end up with constraints well accordance with our crude estimates given in~\eq{masslessg} and~\eq{massiveg}:
\bea{1ex1m0}
g < 6.4 \cdot 10^{-3} \ (\frac{ (E_1/{\rm PeV}) }{ D/(500 \ {\rm Mpc}) } )^{1/4} 
\eea
in the case $\sqrt{s} \gg M_X$ and 
\bea{1ex2m}
g/(M_X/{\rm MeV}) < 0.013 \ (\frac{ 1 }{ (E_1/{\rm PeV})  (D/500 \ {\rm Mpc}) })^{1/4} 
\eea
in the case $\sqrt{s} \ll M_X$. If the secret interactions do exist, then depending on the mass of the mediator either the low energy ($\sqrt{s} \gg M_X$) or the high energy ($\sqrt{s} \ll M_X$) part of the AGN neutrino spectrum will be influenced. 
\\


\noindent{\it 5. Comparison with other constraints.} 
Let us make a comparison of our constraints with the bounds previously obtained from other phenomena. In references~\cite{Bardin},~\cite{Cable},~\cite{Pang} upper bounds to secret 
neutrino-neutrino interactions have been derived by using light meson decays. 
In the LEP constraint (\ref{FVlimit}), based on the measurement of the invisible width of the $Z$ boson, the mediator of the secret interaction is assumed to be a vector particle much heavier than the $Z$ boson, i.e. $M_V\gg M_Z$. In order to compare this bound with the one we obtained, we must extrapolate the LEP constraint to lower mediator masses. 
The coupling constant is of the form $F_V = g^2/M_V^2 < 400G_F$,
and if we set $M_V=100 \ {\rm GeV}$, we will get $g^2 <  4.0 \times 10^{6} G_F$.
A comparison with our result (\ref{1ex2m}) with representative values $E_1 = 1 \ {\rm Pev}$ and $D=500 \ {\rm Mpc}$, giving  $g/M_X <  13/{\rm GeV}$, shows that our constraint is more stringent whenever  
$M_X < 0.5 \ {\rm GeV}$. 

From supernova SN1987A data two kinds of limits were obtained to the neutrino-neutrino couplings, considered in~\cite{Manohar} and~\cite{Kolb}. In~\cite{Manohar} the constraint was derived from the duration of the neutrino pulse and the magnitude of neutrino flux. A limit to the neutrino-neutrino cross section was placed to $\sigma_{\bar\nu \bar\nu} < 10^{-35} \ {\rm cm}^2$, and $g^2 < 5.1\cdot 10^{-6}(T_{\nu}/10 \ {\rm MeV})$.\footnote{According to reference~\cite{Dicus} this result is disputable and the upper limit could actually be somewhat larger.}  It was assumed that the possible secret interaction was due to the exchange of a scalar particle with mass smaller than the neutrino temperature $T_{\nu}$, which is less than $10 \ {\rm MeV}$. Our result is of the same order of magnitude when this limit is interpreted in terms of vector particle exchange. 

In~\cite{Kolb} the supernova neutrino interactions with cosmic neutrino background was used to derive limits for the couplings. The energy of the supernova neutrinos is around $10 \ {\rm MeV}$ and the distance 55 kpc is much shorter than that of AGN. The limits obtained are comparable with those we obtained, $g<5.6 \cdot 10^{-4}$ if $M_V \ll 60 \ {\rm eV}$ and $g/(M_V/{\rm MeV}) < 12$ if $M_V \gg 60 \ {\rm eV}$ (in this case $\sqrt{s}\simeq 60 \ {\rm eV}$). Comparing these results with our equations (\ref{1ex1m0}) and (\ref{1ex2m}) shows that the supernova constraints are more stringent than the AGN constraints in the case the mass of the mediator is below 1 keV (assuming again $E_1 = 1 \ {\rm PeV}$ and $D=500 \ {\rm Mpc}$). With larger mass values the AGN limits always surpass the supernova ones. For example with $M_X = 10 \ {\rm keV}$ we obtain $g<6.4\cdot 10^{-3}$ whereas the supernova limit is $g<0.12$.

The nucleosynthesis limit was obtained~\cite{Masso} by requiring that right-handed Dirac neutrinos have to decouple at temperatures above QCD phase transition. This leads to an upper limit for the coupling strength which in the case of massive mediator is obtained to be $F_V < 3\times 10^{-3} G_F$ and in the case of massless mediator $g<2\times 10^{-5}$. These limits are actually more stringent than any other limit discussed above, including ours, but they are, however, valid only for a very limited class of models.
\\

\noindent{\it 6. Summary.} We have considered a testing of non-standard neutrino-neutrino interactions by using ultrahigh energy AGN neutrinos. If such interactions exist, AGN neutrino flux should be influenced by extra collisions with cosmic background neutrinos. We have derived the upper limit for the coupling strength one can probe by measuring the AGN neutrino flux. Depending on the mass of the mediator of these secret interactions the constraints we obtained, given in equations (\ref{1ex1m0}) and (\ref{1ex2m}), are more stringent than previously presented astrophysical and laboratory limits. 
\\

\noindent{\it Acknowledgements.} I would like to thank Jukka Maalampi and Kimmo Kainulainen for useful discussions and Jukka Maalampi also for a careful reading of the manuscript. I am also indebted to the Jenny and Antti Wihuri foundation for financial support.



\end{document}